\documentclass[pdflatex,sn-mathphys-num]{sn-jnl}

\usepackage{graphicx}%
\usepackage{multirow}%
\usepackage{longtable}
\usepackage{amsmath,amssymb,amsfonts}%
\usepackage{amsthm}%
\usepackage{mathrsfs}%
\usepackage[title]{appendix}%
\usepackage{xcolor}%
\usepackage{textcomp}%
\usepackage{manyfoot}%
\usepackage{acronym}
\usepackage{booktabs, multirow, makecell}
\usepackage{booktabs}%
\usepackage{algorithm}%
\usepackage{algorithmicx}%
\usepackage{algpseudocode}%
\usepackage{listings}%
\usepackage{orcidlink}
\newtheorem{assumption}{Assumption}

\theoremstyle{thmstyleone}%
\theoremstyle{thmstyletwo}%

\theoremstyle{thmstylethree}%

\begin{document}

\title[Article Title]{RL-Aided Cognitive ISAC: Robust Detection and Sensing–Communication Trade-offs}


\author*[1]{\fnm{Adam} \sur{Umra} \orcidlink{0009-0004-2071-2897}}\email{adam.umra@rub.de}

\author[2]{\fnm{Aya} \sur{M. Ahmed} \orcidlink{0000-0002-9698-9057}}\email{aya.mostafaibrahimahmad@rub.de}

\author[1]{\fnm{Aydin} \sur{Sezgin} \orcidlink{0000-0003-3511-2662}}\email{aydin.sezgin@rub.de}

\affil[1]{\orgname{Ruhr University Bochum}, \country{Germany}}

\affil[2]{ \orgname{Robert Bosch GmbH}, \country{Germany}}

\abstract{This paper proposes a reinforcement learning (RL)-aided cognitive framework for massive MIMO-based integrated sensing and communication (ISAC) systems employing a uniform planar array (UPA). The focus is on enhancing radar sensing performance in environments with unknown and dynamic disturbance characteristics. A Wald-type detector is employed for robust target detection under non-Gaussian clutter, while a SARSA-based RL algorithm enables adaptive estimation of target positions without prior environmental knowledge. Based on the RL-derived sensing information, a joint waveform optimization strategy is formulated to balance radar sensing accuracy and downlink communication throughput. The resulting design provides an adaptive trade-off between detection performance and achievable sum rate through an analytically derived closed-form solution. Monte Carlo simulations demonstrate that the proposed cognitive ISAC framework achieves significantly improved detection probability compared to orthogonal and non-learning adaptive baselines, while maintaining competitive communication performance. These results underline the potential of RL-assisted sensing for robust and spectrum-efficient ISAC in next-generation wireless networks.}



\maketitle
\section{Introduction}\label{sec1}
Integrated Sensing and Communication (ISAC)~\cite{wei2023ISAC} has emerged as a central paradigm for sixth-generation (6G) wireless systems~\cite{alsabah20216G}, unifying sensing and communication within a shared framework of spectrum, signal processing, and hardware resources. By breaking the traditional separation between these two functionalities, ISAC not only improves spectral efficiency but also enables wireless networks to perceive, interpret, and interact intelligently with their physical environment. This dual capability supports applications such as autonomous transport, smart cities, industrial automation, and next-generation wireless monitoring~\cite{liu2022ISAC}. Within this broader vision, radar technology serves as a cornerstone by providing high-resolution environmental awareness. When radar sensing is jointly designed with communication links, the resulting system can achieve real-time situational understanding, enhance network intelligence, and facilitate informed decision-making across distributed and autonomous infrastructures~\cite{sturm2011Joint}. 

A promising research direction within advanced radar systems is \emph{Cognitive Radar} (CR), which extends conventional radar paradigms by embedding intelligence, adaptability, and environmental awareness into the sensing process~\cite{haykin2012CR, haykin2006CR}. Unlike traditional radars that operate under fixed transmission parameters, CR employs the perception–action cycle, wherein the system continuously senses the environment, interprets received observations (e.g., target detections, clutter, or interference), and adaptively adjusts its transmission and reception strategies. This closed-loop interaction between transmitter and receiver enables real-time optimization of waveform design, pulse repetition frequency, and beamforming direction. Consequently, CR demonstrates superior detection and tracking performance in dynamic, cluttered, and interference-dominated environments, where static radar configurations often fail to maintain reliability~\cite{greco2016CR,gurbuz2019CR,guerci2010CR}. Furthermore, CR represents a step toward self-optimizing sensing systems that can autonomously adapt to environmental uncertainty without human intervention.

Building upon these adaptive principles, \emph{Reinforcement Learning} (RL) has recently gained attention as a powerful enabler for achieving fully autonomous radar operation~\cite{ahmed2021RL}. RL provides a data-driven framework in which the radar learns optimal sensing and transmission policies through interaction with its environment. By optimizing a reward function aligned with radar performance objectives, such as target detection accuracy, interference mitigation, and energy efficiency, the RL agent iteratively refines its decision-making process through a balance between exploration and exploitation~\cite{sutton2018RL}. This paradigm eliminates the need for prior statistical knowledge of the environment, making RL particularly suitable for cognitive radar applications characterized by nonstationary and uncertain conditions. Recent studies demonstrate that RL-driven beamforming strategies for massive MIMO CR systems can achieve robust multi-target detection without explicit disturbance modeling, outperforming both conventional and adaptive non-learning baselines, particularly in low signal-to-noise ratio (SNR) and dynamically varying environments~\cite{wang2018RL,ahmed2021RL,lisi2022RL,weitong2022RL,wang2024RL,umra2025RL}.

Extending RL methodologies to the ISAC domain, recent research has investigated joint waveform design approaches that simultaneously meet both sensing and communication requirements within a unified signal framework. Unlike conventional techniques that sequentially adapt radar or communication waveforms, these approaches aim to co-optimize both functionalities from a shared resource perspective~\cite{liu2018DFRC}. For instance, in~\cite{ahmed2023ISAC}, an RL-based framework for joint waveform optimization in cognitive massive MIMO ISAC systems was proposed, employing a uniform linear array (ULA) ISAC MIMO base station (BS) to enable efficient spectrum sharing without prior environmental knowledge. By iteratively probing the surroundings, updating the waveform, and balancing sensing and communication objectives, the system achieved enhanced target detection accuracy while preserving communication performance. Simulation results demonstrated consistent superiority over traditional orthogonal waveform designs, particularly under low-SNR conditions. However, this study did not explore the influence of varying communication user density on sensing performance, nor did it analyze the impact of different trade-off parameters that prioritize either sensing or communication objectives. Moreover, the analysis was limited to a ULA-based BS configuration, which restricts spatial adaptability.

To address these limitations, this paper proposes an enhanced cognitive ISAC framework employing a two-dimensional uniform planar array (UPA), extending our previous work~\cite{umra2025RL}. In contrast to existing cognitive ISAC studies the proposed design jointly exploits spatial diversity in both azimuth and elevation to enhance adaptive beamforming and sensing resolution. A rigorous mathematical formulation of the ISAC system is developed, wherein a Wald-type detector~\cite{fortunati2020Wald} is utilized to achieve robust target detection in massive MIMO systems operating under uncertain and dynamic environments. Building on this formulation, an RL-based optimization strategy is introduced to enable adaptive detection and waveform design in scenarios with limited prior knowledge of the environment. Based on the optimized sensing waveform, a trade-off waveform is further constructed to balance sensing accuracy and communication throughput effectively. Extensive Monte Carlo simulations are performed to evaluate the proposed framework, demonstrating that the cognitive ISAC design consistently outperforms conventional benchmark approaches, including orthogonal waveform schemes and adaptive non-RL (NRL) methods, by achieving superior sensing performance while maintaining competitive communication rates.

The remainder of this paper is organized as follows. Section~\ref{sec2} presents the detailed system model, outlines the problem formulation, and describes the proposed cognitive ISAC framework, including the RL-based decision-making mechanism. Section~\ref{sec3} provides a comprehensive analysis of simulation results, emphasizing the performance trade-offs between sensing accuracy and communication quality across different operational conditions, including varying SNR levels and target mobility patterns. Finally, Section~\ref{sec4} concludes the paper by summarizing the key findings, highlighting the contributions in the context of state-of-the-art ISAC research.

\section{Methods}\label{sec2}
In this section, we provide a comprehensive description of the methodologies adopted in this work. We begin by formulating the integrated sensing and communication (ISAC) system model, detailing the signal structures for both radar sensing and data transmission functionalities. Subsequently, we introduce the proposed cognitive ISAC framework, which enables the joint optimization of waveform parameters to achieve an adaptive balance between sensing accuracy and communication reliability. The formulation integrates both domains under a unified optimization criterion, ensuring efficient resource utilization and mutual coexistence. Finally, we describe the RL–based adaptation mechanism developed to enhance target detection performance under dynamic operating conditions.

\subsection{System Model}
\begin{figure}[htbp!]
    \centering
    \includegraphics[width=0.6\textwidth]{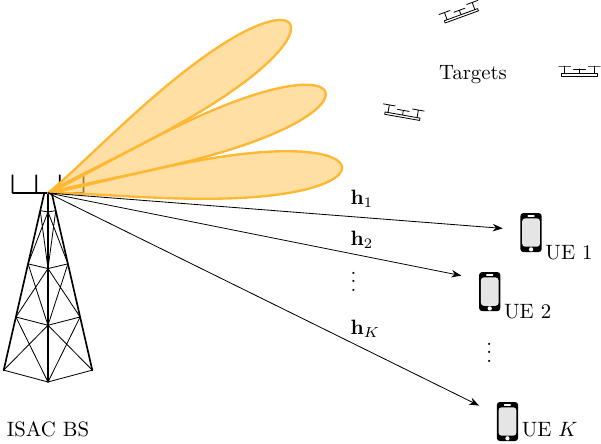}
    \caption{ISAC BS system serving $K$ single-antenna users while operating as a monostatic cognitive radar.}
    \label{fig:ISACsys}
\end{figure}

We consider a MIMO ISAC base station (BS) that simultaneously performs dual functionalities of wireless communication and radar sensing, as depicted in Fig.~\ref{fig:ISACsys}. Specifically, the BS transmits information-bearing symbols to $K$ single-antenna downlink users, while concurrently operating as a monostatic cognitive radar. In this configuration, a unified transmit waveform is employed for both tasks: conveying data to the communication users and probing the surrounding environment for target detection. This joint utilization of the transmitted signal enables efficient spectrum sharing and hardware reuse, thereby enhancing the overall system performance and resource efficiency.

\subsubsection{Radar Signal Model}
Consider a colocated MIMO radar equipped with $N_t$ transmit and $N_r$ receive antennas, yielding a total of $N = N_t N_r$ virtual channels. Both arrays are modeled as UPAs with half-wavelength element spacing~\cite{umra2025RL}. The target is located at elevation angle $\theta$ and azimuth angle $\phi$.

At the $l$-th signaling interval ($l = 1, \ldots, L$), the transmitted narrowband waveform vector $\mathbf{x}_l \in \mathbb{C}^{N_t}$ is emitted, where $L$ denotes the code length~\cite{wu2023Wave}. The corresponding received baseband echo is given by
\begin{equation}
\mathbf{y}_l = \alpha \, \mathbf{a}_r(\theta,\phi) \, \mathbf{a}_t(\theta,\phi)^T \mathbf{x}_l + \mathbf{c}_l ,
\end{equation}
where $\mathbf{y}_l \in \mathbb{C}^{N_r}$ is the received signal vector, $\alpha \in \mathbb{C}$ represents the target’s complex scattering coefficient accounting for the radar cross section and two-way propagation loss~\cite{swerling1960Prob}, and $\mathbf{c}_l$ denotes the additive clutter-plus-noise term, later modeled as a autoregression (AR) process with covariance $\mathbf{\Gamma}$.

The transmit and receive steering vectors $\mathbf{a}_t(\theta,\phi) \in \mathbb{C}^{N_t}$ and $\mathbf{a}_r(\theta,\phi) \in \mathbb{C}^{N_r}$ follow the separable UPA structure~\cite{tan2017UPA}
\begin{equation}
\mathbf{a}(\theta,\phi) = \mathbf{a}_x(\theta,\phi) \otimes \mathbf{a}_y(\theta,\phi),
\end{equation}
where $\otimes$ denotes the Kronecker product, and $\mathbf{a}_x(\theta,\phi)$ and $\mathbf{a}_y(\theta,\phi)$ represent the steering responses along the horizontal and vertical array dimensions, respectively.

By stacking the $L$ received snapshots into a matrix form, the overall signal model becomes
\begin{equation}\label{eq:stacked}
\mathbf{Y}_s = \alpha \, \mathbf{a}_r(\theta,\phi) \mathbf{a}_t(\theta,\phi)^T \mathbf{X} + \mathbf{C}_s ,
\end{equation}
where $\mathbf{X} = [\mathbf{x}_1, \ldots, \mathbf{x}_L] \in \mathbb{C}^{N_t \times L}$ contains the transmitted signals, and $\mathbf{Y}_s, \mathbf{C}_s \in \mathbb{C}^{N_r \times L}$ denote the received and disturbance matrices, respectively.

After correlating with the transmitted waveform matrix at the receiver and vectorizing the result,~\eqref{eq:stacked} can be expressed in terms of the transmit covariance matrix $\mathbf{R} = \tfrac{1}{L}\mathbf{X}\mathbf{X}^H$~\cite{ahmed2014MIMO} as
\begin{equation}
\mathbf{y}_s = \alpha \left[ \mathbf{a}_r(\theta,\phi) \otimes \mathbf{a}_\mathrm{t}(\theta,\phi)^T \mathbf{R} \right] + \mathbf{c}_s ,
\end{equation}  
where $\mathbf{y}_s, \mathbf{c}_s \in \mathbb{C}^{N}$ are the vectorized received signal and disturbance terms, respectively, and $\mathbf{R} \in \mathbb{C}^{N_t \times N_t}$ is the transmit covariance matrix. Defining
\begin{equation}
\mathbf{h} = \mathbf{a}_r(\theta,\phi) \otimes \mathbf{a}_\mathrm{t}(\theta,\phi)^T \mathbf{R} \in \mathbb{C}^N,
\end{equation}
the model reduces to the compact form
\begin{equation}
\mathbf{y}_s = \alpha \mathbf{h} + \mathbf{c}_s .
\end{equation}

To enable spatial processing, the radar field-of-view is discretized into $L_x \times L_y$ angular bins, indexed by spatial frequencies
\begin{equation}
\nu_x^{(i)}, \quad i = 1, \dots, L_x, \qquad \nu_y^{(j)}, \quad j = 1, \dots, L_y ,
\end{equation}
with
\begin{equation}
\nu_x = \tfrac{1}{2}\sin(\theta)\cos(\phi), \qquad \nu_y = \tfrac{1}{2}\sin(\theta)\sin(\phi),
\end{equation}
assuming half-wavelength inter-element spacing. Each bin $(i,j)$ corresponds to a unique spatial location $m \in \mathcal{M}$, where $|\mathcal{M}| = L_x L_y$.

Over $P$ coherent pulses indexed by $p = 1,\dots,P$, the signal model for the $m$-th spatial bin is given by
\begin{equation}
\mathbf{y}_{p,m} = \alpha_{p,m} \mathbf{h}_{p,m} + \mathbf{c}_{p,m},
\end{equation}
where $\alpha_{p,m}$ denotes the complex target amplitude and $\mathbf{c}_{p,m}$ represents clutter-plus-noise at that bin and pulse.

The binary detection problem for each spatial bin and pulse can then be formulated as
\begin{align}
\mathcal{H}_0 &: \; \mathbf{y}_{p,m} = \mathbf{c}_{p,m}, \\
\mathcal{H}_1 &: \; \mathbf{y}_{p,m} = \alpha_{p,m}\mathbf{h}_{p,m} + \mathbf{c}_{p,m}.
\end{align}
A decision statistic $\Lambda_{p,m}$ is compared to a threshold $\eta$ according to
\begin{equation}\label{eq:hyp}
\Lambda_{p,m} \underset{\mathcal{H}_0}{\overset{\mathcal{H}_1}{\gtrless}} \eta ,
\end{equation}
where $\eta = -2 \ln(P_{FA})$ ensures a prescribed false-alarm probability. Employing a Wald-type detector~\cite{fortunati2020Wald}, the test statistic is defined as
\begin{equation}
\Lambda_{p,m} = \frac{2|\mathbf{h}_{p,m}^H \mathbf{y}_{p,m}|^2}{\mathbf{h}_{p,m}^H \hat{\mathbf{\Gamma}}_{p,m} \mathbf{h}_{p,m}},
\end{equation}
where $\hat{\mathbf{\Gamma}}_{p,m}$ denotes the estimated clutter-plus-noise covariance matrix~\cite{ahmed2021RL}. Further details on the estimation of $\mathbf{\Gamma}_{p,m}$ are provided in Appendix~\ref{secA1}. Unlike conventional space–time adaptive processing (STAP) filter, which requires a large set of homogeneous secondary data for clutter covariance estimation, the adopted Wald-type detector operates reliably under limited or non-homogeneous observations. This property is essential in cognitive ISAC scenarios, where the clutter statistics may vary rapidly and only a single snapshot per pulse is available. Moreover, the Wald detector exhibits robustness to non-Gaussian and heavy-tailed disturbances and enables analytical tractability for RL-based adaptation.

\subsubsection{Communication Signal Model}
The dual-functional radar-communication system must also guarantee reliable downlink performance. Specifically, the transmitted signal $\mathbf{X}$ simultaneously serves as both a radar probing waveform and a communication data carrier~\cite{liu2018DFRC}. This inherent coupling necessitates a careful joint design to satisfy radar beam pattern requirements while maintaining communication quality-of-service (QoS) criteria, such as the achievable sum-rate and symbol error rate (SER)~\cite{liu2018DFRC}.

The received downlink signal at $K$ single-antenna users can be expressed as  
\begin{equation}\label{eq:commsig}
\mathbf{Y}_c = \mathbf{H}\mathbf{X} + \mathbf{N},
\end{equation}  
where $\mathbf{H} \in \mathbb{C}^{K \times N_t}$ denotes the flat Rayleigh fading channel matrix, $\mathbf{X} \in \mathbb{C}^{N_t \times L}$ represents the dual-use transmit waveform, and $\mathbf{N} \in \mathbb{C}^{K \times L}$ is additive white Gaussian noise (AWGN). The channel $\mathbf{H}$ is assumed to be perfectly known at the transmitter and to remain constant within one radar pulse or one communication frame.  

Let $\mathbf{S} \in \mathbb{C}^{K \times L}$ denote the target constellation matrix corresponding to the intended information symbols for all users. The received signal can equivalently be decomposed as  
\begin{equation}
\mathbf{Y}_c = \mathbf{S} + \underbrace{\left(\mathbf{H}\mathbf{X} - \mathbf{S}\right)}_{\text{MUI}} + \mathbf{N},
\end{equation}  
where the underbraced term quantifies the multi-user interference (MUI) resulting from imperfect spatial separation among users. The overall MUI energy is defined as  
\begin{equation}
P_{\text{MUI}} = \|\mathbf{H}\mathbf{X} - \mathbf{S}\|_F^2,
\end{equation}  
which directly influences the achievable downlink throughput and user fairness.  

For the $k$-th user, the instantaneous signal-to-interference-plus-noise ratio (SINR) per frame is given by  
\begin{equation}
\gamma_k = \frac{\mathbb{E}\left[|s_{k,j}|^2\right]}{\mathbb{E}\left[|h_k^T x_j - s_{k,j}|^2\right] + N_0},
\end{equation}  
where $s_{k,j}$ denotes the $(k,j)$-th entry of $\mathbf{S}$, $h_k^T$ is the $k$-th row of $\mathbf{H}$, and $x_j$ is the $j$-th column of $\mathbf{X}$. Consequently, the overall achievable downlink sum-rate can be expressed as  
\begin{equation}
R = \sum_{k=1}^{K} \log_2 \left(1 + \gamma_k\right).
\end{equation}  

\subsection{Cognitive ISAC for Sensing–Communication Trade-offs}

The performance objectives of communication and radar sensing systems are fundamentally different, reflecting their distinct operational goals. In communication systems, the primary objective is to transmit information reliably and efficiently from the source to the intended receiver. Conversely, radar sensing is not concerned with maximizing information transfer, but with extracting relevant information from the echoes of the transmitted signals. In essence, communication systems embed the information of interest within the transmitted waveform itself, whereas in radar systems, the desired information is contained within the received reflections.

This fundamental distinction motivates the design challenge in ISAC systems: a single waveform must serve two purposes simultaneously. From the system models previously introduced, one approach is to formulate waveform design as an optimization problem that balances dual objectives: (i) minimizing the MUI power for reliable communication, and (ii) achieving a desired beampattern for effective radar detection.  

A widely studied formulation for the joint radar-communication trade-off was introduced in~\cite{liu2018DFRC}, and can be expressed as  
\begin{subequations}\label{eq:P0}
\begin{align}
\mathcal{P}_0:\quad &\min_{\mathbf{X}} \quad \rho \lVert \mathbf{H}\mathbf{X}-\mathbf{S} \rVert_F^2 
+ (1-\rho)\, \lVert \mathbf{X}-\mathbf{X}_0 \rVert_F^2 \tag{\ref{eq:P0}} \\
&\text{s.t.}\quad \frac{1}{L}\lVert \mathbf{X} \rVert_F^2 = P_\mathrm{T}.
\end{align}
\end{subequations}
 The parameter $0 \leq \rho \leq 1$ acts as a trade-off factor that adjusts the priority between the two system objectives. The first term in~\eqref{eq:P0} penalizes MUI, thereby improving communication reliability, while the second term encourages similarity to a reference waveform $\mathbf{X}_0$, thus preserving radar functionality.

To facilitate analysis, the composite objective function can be reformulated as
\begin{align}
\label{eq:reform}
\rho \, \lVert \mathbf{H}\mathbf{X}-\mathbf{S} \rVert_F^2 
+ (1-\rho)\, \lVert \mathbf{X}-\mathbf{X}_0 \rVert_F^2
= \left\lVert \mathbf{A}\mathbf{X}-\mathbf{B}\right\rVert_F^2,
\end{align}
where
\begin{align}
\mathbf{A} &= \begin{bmatrix}
\sqrt{\rho}\,\mathbf{H} \\[3pt]
\sqrt{1-\rho}\,\mathbf{I}_{N_\mathrm{t}}
\end{bmatrix}, 
&\quad
\mathbf{B} &= \begin{bmatrix}
\sqrt{\rho}\,\mathbf{S} \\[3pt]
\sqrt{1-\rho}\,\mathbf{X}_0
\end{bmatrix}.
\end{align}

By applying the reformulation in~\eqref{eq:reform}, the optimization task can be expressed as
\begin{align}
\label{eq:P11}
\mathcal{P}_{0.1}:\quad &\min_{\mathbf{X}}\;\; 
\lVert \mathbf{A}\mathbf{X}-\mathbf{B} \rVert_F^2 \\ \nonumber
&\text{s.t.}\quad \lVert \mathbf{X} \rVert_F^2 = L P_\mathrm{T}.
\end{align}
Problem~$\mathcal{P}_{0.1}$ belongs to the class of quadratically constrained quadratic programs (QCQPs). Since the quadratic equality constraint renders the problem non-convex, solving it exactly is computationally challenging. A widely adopted approach to address this difficulty is to employ the semidefinite relaxation (SDR) framework, which reformulates the QCQP into a semidefinite program (SDP). This convex relaxation allows the use of efficient numerical solvers while still providing practically useful solutions. The optimal solution for this problem is derived in~\cite{liu2018DFRC} as
\begin{equation}
    \mathbf{X}_{opt} = \left(\mathbf{A}^H\mathbf{A}+\lambda_{opt}\mathbf{I}_{N_\mathrm{t}}\right)^\dagger \left(\mathbf{A}^H\mathbf{B}\right),
\end{equation}
where $\lambda_{opt}$ is the optimal eigenvalue of $\mathbf{Q}=\mathbf{A}^H\mathbf{A}$, which can be obtained by simple line search
methods, e.g., Golden-section search. More details on how to derive this solution and the algorithm can be found in Appendix~\ref{appB}.

Having derived the optimal waveform for the considered ISAC system, it remains necessary to construct a reference waveform, denoted as $\mathbf{X}_0$, which is optimal for target detection. To this end, we formulate the following optimization problem  
\begin{align}
\label{eq:P1}
\mathcal{P}_1:\quad & \min_{\mathbf{X}_0}\;\; \lVert \mathbf{H}\mathbf{X}_0 - \mathbf{S} \rVert_F^2 \\ \nonumber
& \text{s.t.} \quad \frac{1}{L}\, \mathbf{X}_0 \mathbf{X}_0^{\mathrm{H}} = \mathbf{R}_d ,
\end{align}
where $\mathbf{R}_d$ represents the desired transmit covariance matrix, which corresponds to a target beampattern. The objective function minimizes the MUI while ensuring that the resulting waveform adheres to the specified covariance constraint $\mathbf{R}_d$.  

A closed-form solution to problem $\mathcal{P}_1$ is provided in~\cite{liu2018DFRC} as  
\begin{equation}
    \mathbf{X}_{0,\mathrm{opt}} = \sqrt{L}\,\mathbf{F}\mathbf{U}\mathbf{I}_{N\times L}\mathbf{V}^{\mathrm{H}},
\end{equation}
where $\mathbf{R}_d = \mathbf{F}\mathbf{F}^{\mathrm{H}}$ and the singular value decomposition (SVD) $\mathbf{U}\mathbf{\Sigma}\mathbf{V}^{\mathrm{H}} = \mathbf{F}^{\mathrm{H}}\mathbf{H}^{\mathrm{H}}\mathbf{S}$ holds. The detailed derivation of this closed-form solution is provided in Appendix~\ref{appC}.

In~\cite{liu2018DFRC}, the optimization problem in $\mathcal{P}_{1}$ was addressed under the assumption that the transmit covariance matrix $\mathbf{R}_d$ is perfectly known a priori. However, this assumption is unrealistic, since both the number of targets and their directions of arrival are typically unknown prior to the sensing process. In this work, we employ RL to estimate the target positions. Based on these estimates, the transmit covariance matrix $\mathbf{R}_d$ is subsequently designed by solving the following optimization problem~\cite{stoica2007Opt}
\begin{align}
\label{eq:P2}
\mathcal{P}_2:\quad & \max_{\mathbf{R}_d} \; \mathrm{tr}(\mathbf{R}_d \hat{\mathbf{B}}) \\
\text{s.t.} \quad & \mathrm{tr}(\mathbf{R}_d) = P_\mathrm{T}, \\
& \mathbf{R}_d \succeq 0,
\end{align}
where 
\begin{equation}
\hat{\mathbf{B}} = \sum_{b=1}^{\hat{b}} \mathbf{a}_\mathrm{t}(\theta_b,\phi_b)\mathbf{a}_\mathrm{t}^{\mathrm{H}}(\theta_b,\phi_b).
\end{equation}
where $\hat{b}$ denotes the number of targets and $\mathbf{a}_\mathrm{t}(\theta_b,\phi_b)$ steering vector of the target $b$.
The closed-form solution of problem $\mathcal{P}_2$ is given by
\begin{equation}
    \mathbf{R}_{d,\mathrm{opt}} = P_\mathrm{T}\,\mathbf{u}\mathbf{u}^{*},
\end{equation}
where $\mathbf{u}$ denotes the unit-norm eigenvector of $\hat{\mathbf{B}}$ corresponding to its largest eigenvalue $\lambda_{\max}(\hat{\mathbf{B}})$~\cite{stoica2007Opt, stoica2002MIMO}. Hence, problem $\mathcal{P}_2$ maximizes the total transmitted power over the estimated target locations. 

Naturally, the positions of the targets are not known \textit{a priori}. Therefore, a RL framework is developed in the following to learn the target locations and calculate the transmit covariance matrix with $\mathcal{P}_2$.

\subsection{Reinforcement Learning Framework}
\begin{figure}[htbp!]
    \centering
    \includegraphics[width=0.8\textwidth]{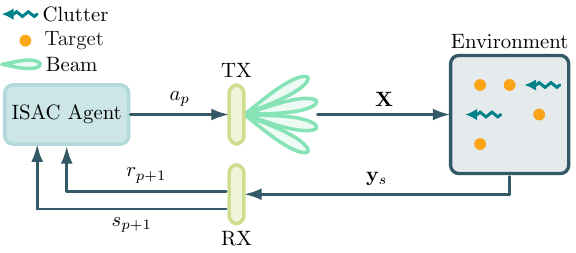}
    \caption{RL framework for the SARSA-based cognitive ISAC.}
    \label{fig:ISACsysRL}
\end{figure}
RL is a branch of machine learning in which an intelligent agent gradually improves its performance by interacting with an environment~\cite{sutton2018RL}. Unlike supervised learning, where the correct output is provided for each example, in RL the agent must discover successful strategies on its own through repeated trials and feedback. This process can be conceptualized as an iterative cycle of trial and error: the agent takes an action, observes how the environment responds, and then adjusts its future behavior in order to achieve better outcomes. Over time, this iterative process enables the agent to learn a policy $\pi$ that is, a set of rules that guide which action should be taken in a given situation. The overall RL mechanism applied in our ISAC scenario is illustrated in Fig.~\ref{fig:ISACsysRL}. In the context of our work, the ISAC system plays the role of the agent. The ISAC agent receives continuous feedback from the surrounding environment, evaluates this feedback, and then decides how to act next. To formalize the learning mechanism, we employ the SARSA algorithm, which is a widely used RL method~\cite{sutton2018RL}. The name ``SARSA'' comes from the sequence of elements involved in each learning step: \emph{State, Action, Reward, next State, and next Action}. At each step, the agent observes its current state $s_p$, chooses an action $a_p$, and then receives a reward $r_{p+1}$ from the environment. After transitioning to a new state $s_{p+1}$, the agent selects its next action $a_{p+1}$. This entire sequence is then used to update the agent’s knowledge of the environment.

The agent’s decision-making is guided by a function known as the Q-value, which measures the long-term benefit of taking a particular action in a given state under a policy $\pi$~\cite{sutton2018RL}. Mathematically, the SARSA update rule is expressed as
\begin{equation}\label{eq:qfunc}
    Q(s_p, a_p) \leftarrow  Q(s_p, a_p) +  \alpha \left[ r_{p+1} + \gamma Q(s_{p+1}, a_{p+1}) - Q(s_p, a_p) \right],  
\end{equation}
where $\alpha \in [0,1]$ is the learning rate that controls how strongly new information influences the Q-values, and $\gamma \in [0,1]$ is the discount factor that determines how much importance is given to future rewards compared to immediate ones. 

In our ISAC setup, SARSA is used to efficiently explore the environment and detect the targets under unknown conditions. The knowledge acquired during learning is stored in a state–action value matrix, denoted by $\mathbf{Q}$. Its size depends on the dimensions of the state and action spaces, which will be introduced later. This matrix is continuously updated so that the joint waveform of the ISAC system can adapt dynamically to changes in the environment.

In the following, we define the specific components of the learning process for the ISAC system: the state, the action, reward and the SARSA-based algorithm.

\subsubsection{States and Actions}
The state observed at each pulse $p$ is represented by state $s_p$.  
This state provides a  representation of the sensing environment at time step $p$, encapsulating the detection outcome of the radar sensing process.
In other words, $s_p$ carries information about what the system has detected. Specifically, the state is the total number of range bins whose corresponding test statistics exceed a predefined detection threshold $\eta$, thereby indicating the presence of a possible target. To formalize the detection outcome, we introduce a binary indicator for each bin $m \in \mathcal{M}$. This indicator captures whether or not a detection is declared in bin $m$ at pulse $p$
\begin{equation}
\widehat{\Lambda}_{p,m} =
\begin{cases}
1, & \text{if } \Lambda_{p,m} > \eta, \\
0, & \text{otherwise}.
\end{cases}
\end{equation}
Here, $\widehat{\Lambda}_{p,m}=1$ indicates that the received test statistic $\Lambda_{p,m}$ surpasses the threshold $\eta$, and thus a detection is declared for bin $m$. Conversely, $\widehat{\Lambda}_{p,m}=0$ denotes that no detection is made in that bin.  
Summing over all bins yields the total number of detections at pulse $p$, which is given by
\begin{equation}\label{eq:state}
T_p = \sum_{m \in \mathcal{M}} \widehat{\Lambda}_{p,m},
\end{equation}
where $T_p \in \{0,1,\dots,\tilde{T}\}$ represents the total number of detections at pulse $p$. The parameter $\tilde{T}$ denotes the maximum number of detectable targets, thereby defining the size of the state space.
 In practice, this limitation ensures that the state space remains finite and manageable for the RL agent. For implementation purposes, the state is defined as $s_p = T_p+1$  and the state space is defined as $\mathcal{S} := \{1, \dots, \tilde{T}+1\}$.

The action chosen at step $p$ is represented by $a_p$. The action is chosen from the action space $\mathcal{A}\in \{\Theta_0,\Theta_1,\dots,\Theta_{\tilde{T}}\}$, and each $\Theta_j$ corresponds to the set of indices associated with the top-$j$ values of $\Lambda_{p,m}$.  
Thus, $\Theta_j$ encodes the set of candidate bins that are prioritized for sensing at pulse $p$. Based on the chosen action, the optimization procedure unfolds in multiple stages. First, problem $\mathcal{P}_2$ is solved using the candidate bins specified by $\Theta_p$, yielding the optimal covariance matrix $\mathbf{R}_d$ that governs the sensing process.  
This intermediate result is then employed in problem $\mathcal{P}_1$, where the optimal waveform $X_0$ is determined to minimize MUI in the communication domain.  
Finally, the solution of $\mathcal{P}_1$ is used to optimize problem $\mathcal{P}_0$.  
The final output of this sequential optimization pipeline is an optimized waveform, which balances sensing accuracy and communication quality, and is transmitted in the subsequent pulse.  
In this way, the RL agent continuously adapts its actions across pulses, enabling a dynamic and context-aware allocation of resources between radar sensing and wireless communication.

\subsubsection{Reward}
Having defined the ISAC agent's action and state spaces, we now focus on designing an appropriate reward function that guides its behavior toward accurate target detection. The reward function plays a key role in shaping the agent’s learning objective. For radar sensing, a natural performance measure is the probability of detection $P_{\mathrm{D}}$, balanced against the probability of false alarm $P_{\mathrm{FA}}$. Intuitively, the reward should encourage correct detections in bins containing targets while penalizing false detections in empty bins.

Let $\mathcal{T}$ denote the set of bins that contain a target, and let $\mathcal{N}$ be its complement. Following~\cite{fortunati2020Wald}, the theoretical detection performance at bin $m$ can be characterized by an asymptotic detection probability $\widehat{P}_{\mathrm{D},(p,m)}$, approximated by a first-order Marcum $Q$-function~\cite{nuttall1974qfunction}
\begin{equation}
    \widehat{P}_{\mathrm{D},(p,m)} \stackrel{N \rightarrow \infty}{=} Q_1\!\left(\sqrt{\kappa_{p,m}}, \sqrt{\eta}\right),
\end{equation}
where $\eta$ is the detection threshold, and $\kappa_{p,m}$ is given by
\begin{equation}
  \kappa_{p,m} =
    \frac{2\,|\widehat{\alpha}_{p,m}|^2 \,\|\mathbf{h}_{p,m}\|^4}{\mathbf{h}_{p,m}^H\,\widehat{\mathbf{\Gamma}}_{m}\,\mathbf{h}_{p,m}},
    \quad
    \widehat{\alpha}_{p,m} =
    \frac{\mathbf{h}_{p,m}^H\mathbf{y}_{p,m}}{\|\mathbf{h}_{p,m}\|^2}.
\end{equation}

Based on this, the reward at time step $p$ is defined as
\begin{equation}\label{eq:reward}
    r_{p}
    =
    \sum_{m \in \mathcal{T}} \widehat{P}_{\mathrm{D},(p,m)}
    -
    \sum_{m \in \mathcal{N}} \widehat{P}_{\mathrm{D},(p,m)},
\end{equation}
which encourages the agent to correctly identify targets while penalizing false alarms.

\subsubsection{SARSA-based Algorithm}
In the previous sections, we defined the key components of our RL framework, including states, actions, and the reward function. Building upon these foundations, we now introduce the SARSA-based algorithm. SARSA is a model-free, on-policy RL method known for inherently balancing exploration and exploitation, making it particularly advantageous in real-time scenarios where environmental conditions are often nonstationary~\cite{sutton2018RL}. Moreover, its simplicity in implementation and adaptive nature contribute to robust performance even under dynamic conditions. Central to the SARSA algorithm is the \emph{Q}-function, as defined before. This function approximates the long-term return, defined as the expected cumulative discounted reward, that an agent accrues by taking action $a$ in state $s$ and subsequently following the current policy. The learning mechanism involves iteratively updating the estimate $Q(s,a)$ using the observed tuple $\{s_k,\, a_k,\, r_{k+1},\, s_{k+1},\, a_{k+1}\}$ \cite{sutton2018RL}. 

To achieve robust performance of the SARSA algorithm, an $\varepsilon$-greedy policy is utilized to balance the trade-off between exploration and exploitation. Under this scheme, the agent primarily selects the optimal action based on the current $Q$-values, while occasionally exploring alternative actions with probability $\varepsilon$. This mechanism allows the system to adapt to environmental variations such as fluctuations in the number of detected targets, SNR, and angular positions. In this study, the quasi $\varepsilon$-greedy policy with target recovery, as proposed in~\cite{lisi2022RL}, is adopted. The policy is formally expressed as
\begin{equation}\label{eq:epsilon_greedy}
a_{p+1}^* = 
\begin{cases}
\arg\max\limits_{a \in \mathcal{A}} Q(s_{p},a), & m_{p+1} < m_p, \\[5pt]
\begin{cases}
\text{a random action from } \mathcal{A}', & \text{w.p. } \varepsilon, \\[3pt]
\arg\max\limits_{a \in \mathcal{A}} Q(s_{p+1},a), & \text{w.p. } 1-\varepsilon,
\end{cases} & m_{p+1} \geq m_p,
\end{cases}
\end{equation}
where $\mathcal{A}'$ denotes the subset of actions corresponding to a higher number of angular bins than those in the current state. According to this formulation, when the number of detected targets at time step $p+1$ is equal to or greater than that at step $p$, the radar employs the exploratory strategy defined for the condition $m_{p+1} \geq m_p$. This prevents the system, during exploration, from concentrating its power on fewer angular bins than required for the number of potential targets. Conversely, when the number of detected targets decreases, the policy enforces a recovery phase by selecting the greedy action associated with the previous state $s_p$.

Building on these foundational principles, Algorithm~\ref{alg:alg1} details the complete SARSA-based RL framework for ISAC. The algorithm integrates the SARSA methodology to iteratively optimize the waveform and facilitate a trade-off between robust detection decisions and high communication rates.
\begin{algorithm}[H]
\caption{SARSA-Based RL Algorithm for Cognitve ISAC}\label{alg:alg1}
\textbf{Initialize} $p \leftarrow 0$, $\mathbf{Q}  \leftarrow \mathbf{0}_{(|\mathcal{S}|) \times |\mathcal{A}|)}$, $s_0  \leftarrow 1$, $a_0  \leftarrow 1$, and $\mathbf{X}_0  \leftarrow \mathbf{I}$.

\textbf{repeat} for each time step $p$:
\begin{algorithmic}[1]
    \State Take action $a_p$ by transmitting the waveform $\mathbf{X}_p$
    \State $\mathbf{y}_{p,m}\leftarrow$ acquired received signal $\forall m$
    \State $s_{p+1}\leftarrow$ result from~\eqref{eq:state}, $r_{p+1}\leftarrow$ result in~\eqref{eq:reward} 
    \State $a_{p+1}\leftarrow$ as in~\eqref{eq:epsilon_greedy}, identify $\Theta_{p+1}$
    \State $\mathbf{R}_{d,p+1}\leftarrow$ solution $\mathcal{P}_2$
    \State $\mathbf{X}_{0,p+1}\leftarrow$ solution in $\mathcal{P}_1$
    \State $\mathbf{X}_{p+1}\leftarrow$ solution in $\mathcal{P}_0$
    \State Update $Q(s_k, a_k)$ using ~\eqref{eq:qfunc}, $s_k \leftarrow s_{p+1}$, $a_p \leftarrow a_{p+1}$.
\end{algorithmic}
\textbf{until} the observation period $P$ terminates.
\end{algorithm}

\section{Results and Discussion}\label{sec3}
 The performance of the proposed RL framework is assessed through a series of simulation experiments. The following subsection outlines the simulation setup, including all parameter configurations and the adopted disturbance model. Subsequently, various simulation scenarios are examined and discussed, beginning with stationary targets under different numbers of communication users, and extending to dynamic environments characterized by target appearance, disappearance, and mobility.

\subsection{Parameters}
In our simulations, a massive MIMO BS is considered, which transmits messages of length $L = 30$ to $K$ users over a channel matrix $\mathbf{H}$ with elements $H_{i,j} \in \mathcal{C}\mathcal{N}(0,1)$. The channel is assumed to be perfectly estimated. The transmitted symbols are drawn from a quadrature phase-shift keying (QPSK) constellation with unit average power. The simulations are conducted over a spatial grid defined by $L_x = L_y = 11$, resulting in a total of $121$ bins. The grid is represented in the spatial frequency domain, where $\nu_x$ and $\nu_y$ take values from the discrete set $\{-0.5, -0.4, \dots, 0.5\}$. The key simulation parameters are summarized in Table~\ref{tab:par}. Unless noted otherwise, these parameters remained fixed for all subsequent experiments. The following simulations were each carried out using $10^3$ Monte Carlo runs.

\begin{table}[htbp!]
\caption{Simulation Parameters\label{tab:par}}
\centering
\begin{tabular}{@{}ll@{}}
\toprule
\textbf{Parameter}  & \textbf{Value} \\
\midrule
Probability of false alarm, $P_\mathrm{FA}$ & $10^{-4}$ \\ 
Number of transmit antennas, $N_r$ & $100$ \\
Number of receive antennas, $N_t$ & $100$ \\
Transmit power, $P_T$ & $1$ \\ 
Tail heaviness parameter, $\mu$ & $2$ \\
Noise variance, $\sigma^2_w$ & $1$ \\
Number of detectable targets, $\tilde{T}$ & $10$ \\
Discount factor, $\gamma$ & $0.8$ \\
Exploration factor $\epsilon$ & $0.5$\\
Learning rate, $\eta$ & $0.8$ \\
\botrule
\end{tabular}
\end{table}

The disturbance vector $\mathbf{c}_s$ is modeled as a two-dimensional (2D) circular complex autoregressive (AR) process~\cite{dudgeon1990SP}. The process is defined as
\begin{equation}
    c_{n_x, n_y} = \sum_{i=1}^{p} \sum_{j=1}^{q} \rho_{i,j} \, c_{n_x - i, n_y - j} + w_{n_x, n_y},
\end{equation}
where $c_{n_x, n_y}$ denotes the disturbance component at spatial index $(n_x, n_y)$. The coefficients $\rho_{i,j}$ capture the spatial dependencies of the AR process, thereby characterizing its correlation structure. The term $w_{n_x, n_y}$ represents the driving noise at each spatial location, which serves as a stochastic excitation to the system.

To more accurately model realistic radar clutter, the driving noise $w_{n_x, n_y}$ is assumed to follow a heavy-tailed Student's $t$-distribution~\cite{fortunati2020Wald}, given by
\begin{equation}
    p_w(w_{n_x, n_y}) = \frac{\mu}{\sigma^2_w} 
    \left( \frac{\mu}{\xi} \right)^{\!\mu} 
    \left( \frac{\mu}{\xi} + \frac{|w_{n_x, n_y}|^2}{\sigma^2_w} \right)^{-(\mu + 1)},
\end{equation}
where $\mu$ controls the heaviness of the distribution tails and $\sigma^2_w$ denotes the noise variance. This formulation effectively captures the impulsive and non-Gaussian behavior commonly encountered in high-clutter radar environments, enhancing robustness in practical detection scenarios. The values of $\mu$ and $\sigma^2_w$ used in these simulations are listed in Table~\ref{tab:par}.

The corresponding power spectral density (PSD) of the 2D AR process in the spatial frequency domain is expressed as
\begin{equation}
     S(\nu_x, \nu_y) = \sigma^2_w 
     \left| 1 - \sum_{n=1}^{p}\sum_{l=1}^{q} 
     \rho_{n,l} e^{-j 2 \pi (n \nu_x + l \nu_y)} \right|^{-2},
\end{equation}
which characterizes the distribution of disturbance power across spatial frequencies. The AR coefficients directly influence the spectral shape of the disturbance process, thereby determining its spatial correlation properties. The coefficient matrix employed in these simulations is given by
\begin{equation}
    \boldsymbol{\rho} = 
    \begin{bmatrix} 
        0 & 0.1 & 0.1 \\ 
        0.1 & 0 & 0 \\ 
        0.05 & 0 & 0 
    \end{bmatrix}.
\end{equation}
This matrix defines the spatial correlation structure of the modeled disturbance field.

\subsection{Comparison of RL-Based and Benchmark Schemes in a Stationary Scene}\label{sec:comp_rl}
This section presents an evaluation of the proposed system’s performance in a stationary sensing environment. Within this scenario, four distinct targets are assumed to be present and detected by the sensing module. The spatial locations of these targets, along with their corresponding SNRs are summarized in Table~\ref{tab:scene1}. The communication subsystem is configured to serve $K = 48$ users, a feasible setting given the availability of $N_t = 100$ transmit antennas. The communication link operates at an SNR of 12~dB. To assess the efficacy of the proposed cognitive RL-driven ISAC framework, we compare its performance against two benchmark schemes. The first employs orthogonal beam pattern selection at each time step, while the second represents an adaptive non-RL (NRL) approach, wherein a beam pattern is iteratively generated to focus energy toward the spatial bins that exceed the detection threshold. For this preliminary investigation, we set the weighting factor to $\rho = 0.2$, thereby assigning a lower priority to communication performance relative to sensing accuracy.

\begin{figure}[htbp!]
    \centering
    \includegraphics[width=1\textwidth]{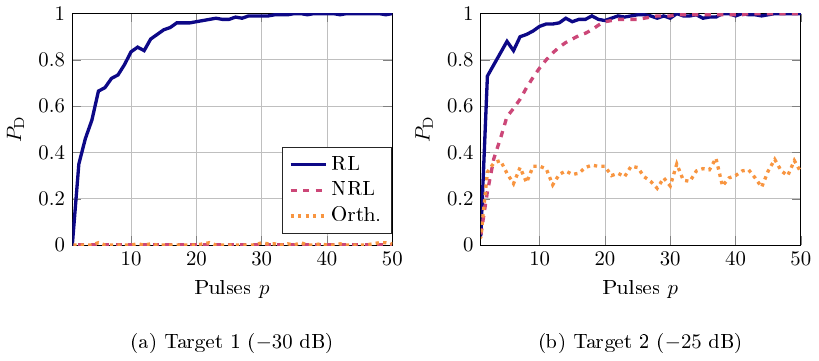}
    \caption{Probability of detection over pulses for Target 1 and 2 and with $\rho = 0.2$ and $K = 48$}
    \label{fig:prob_stat}
\end{figure}

\begin{table}[htbp!]
\centering
\caption{Stationary Targets}
\label{tab:scene1}
\renewcommand{\arraystretch}{1.2} 
\setlength{\tabcolsep}{8pt} 
\begin{tabular}{c c}
\toprule
\textbf{Target} & \multicolumn{1}{c}{\textbf{Time Interval}} \\
\cmidrule(lr){2-2}
 & \textbf{[1, 50]} \\
\midrule
\multirow{2}{*}{1} 
  & (-0.4, -0.4)\footnotemark[1] \\
  & \makecell{$-30$ dB\footnotemark[2]} \\
\midrule
\multirow{2}{*}{2} 
  & (0, 0)  \\
  & $-25$ dB  \\
  \midrule
  \multirow{2}{*}{3} 
  & (0.3, 0.1)   \\
  & $-20$ dB  \\
  \midrule
  \multirow{2}{*}{4} 
  & (-0.1, 0.4)  \\
  & $-15$ dB \\
\bottomrule
\end{tabular}
\footnotetext[1]{Target’s position.}
\footnotetext[2]{Target’s SNR.}
\end{table}
Figure~\ref{fig:prob_stat} illustrates the temporal evolution of the detection probability for Target~1 and~2 across the considered approaches. These two targets are selected for analysis as they exhibit the lowest SNR values and thus represent the most challenging detection cases within the environment. As observed, the proposed RL-based method consistently outperforms both benchmark schemes. Specifically, neither the orthogonal beam pattern nor the NRL approach is able to detect Target~1. While the NRL method succeeds in detecting Target~2, it does this slower compared to the RL framework. The orthogonal beam pattern approach attains a maximum detection probability of only $\sim 0.3$ and lacks adaptability, as it does not incorporate any adaptive mechanism.

\begin{figure}[htbp!]
    \centering
    \includegraphics[width=0.5\textwidth]{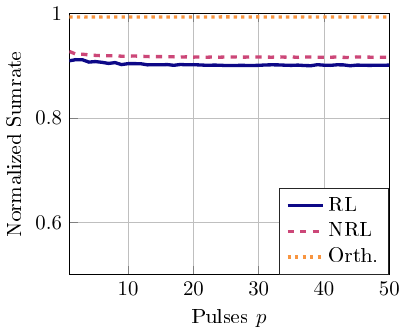}
    \caption{Sum rate over time with $\rho = 0.2$, $K = 48$ and transmit SNR of $12$~dB}
    \label{fig:sum_stat}
\end{figure}

Figure~\ref{fig:sum_stat} illustrates the normalized sum rate with respect to the zero-MUI benchmark over the time. As anticipated, the orthogonal beamforming configuration achieves the highest sum rate, owing to its absence of inter-beam interference. In contrast, both the RL and NRL schemes generate beams directed toward distinct spatial regions, thereby increasing MUI and reducing the overall throughput.
It is further observed that the sum rate remains relatively stable over time across all approaches. However, in the proposed RL-based method, a transient higher sum rate is observed during the initial pulses, corresponding to the learning phase in which the agent explores and adapts to the target locations.



\subsection{Impact of Sensing–Communication Trade-Offs on System Performance}
In the following analysis, the performance of the cognitive ISAC framework is evaluated under various sensing–communication trade-offs. The target positions, SNRs, and the number of communication users remain identical to those defined in Section~\ref{sec:comp_rl}. In the following, we additionally investigate the sum rate performance for different communication SNR levels. 

\begin{figure}[htbp!]
    \centering
    \includegraphics[width=1\textwidth]{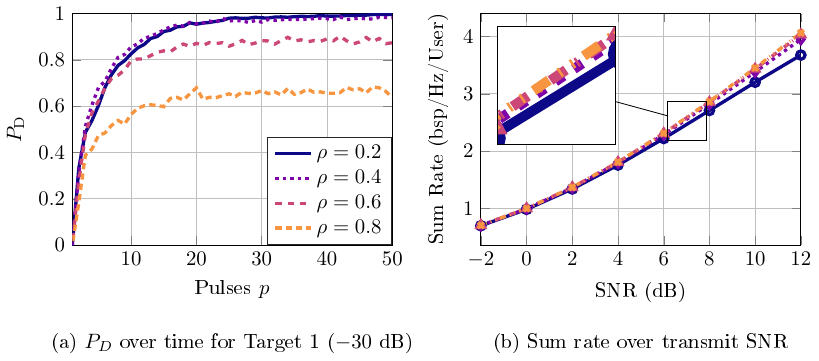}
    \caption{Probability of detection over time and sum rate over varying transmit SNRs for different trade-off $\rho$}
    \label{fig:comp_trade}
\end{figure}

Figure~\ref{fig:comp_trade}~(a) depicts the temporal evolution of the detection probability for different trade-off factors $\rho$. Only Target~1 is considered, as it exhibits the lowest SNR and thus represents the most challenging detection scenario. As expected, smaller values of $\rho$ result in improved detection performance, as they assign greater emphasis to sensing optimization. Notably, for $\rho = 0.2$ and $\rho = 0.4$, the detection probabilities are nearly identical, with $\rho = 0.4$ exhibiting only a marginally lower performance. Figure~\ref{fig:comp_trade}~(b) presents the average per-user sum rate across varying communication SNR levels. For low SNRs, the sum rate remains comparable across all trade-off factors. However, beyond 6~dB, the configuration with $\rho = 0.2$ begins to lag behind the others. The remaining trade-off settings yield similar performance up to approximately 12~dB, where a slight degradation is observed for $\rho = 0.4$. Overall, these results indicate that a trade-off factor of $\rho = 0.4$ provides a balanced compromise, offering strong sensing performance while maintaining a competitive communication sum rate.

\subsection{Effect of the Number of Communication Users on Detection and Sum Rate}
\begin{figure}[htbp!]
    \centering
    \includegraphics[width=1\textwidth]{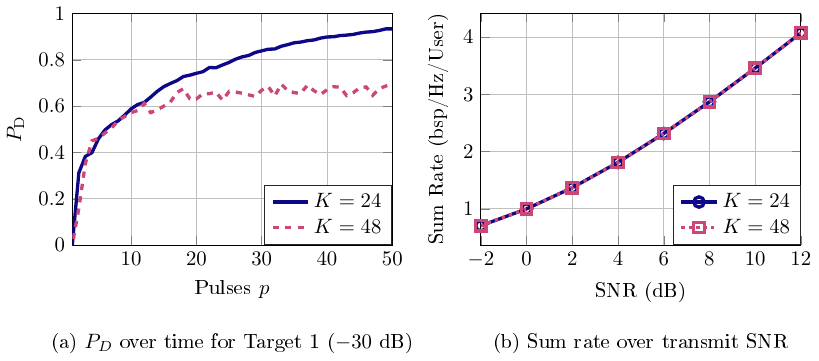}
    \caption{Probability of detection over time and sum rate over varying transmit SNRs for different number of user $K$}
    \label{fig:comp_user}
\end{figure}

Up to this point, the analysis has been conducted assuming $K = 48$ communication users. In the following, we examine how system performance changes when the number of users is reduced. Specifically, we compare the detection probability and the sum rate for both high and low user densities. The target positions and SNR values are again identical to those listed in Table~\ref{tab:scene1}. For the subsequent results, we set the trade-off factor to $\rho = 0.8$, thereby prioritizing communication performance during waveform optimization. As shown in Figure~\ref{fig:comp_user}~(a), a lower number of communication users yields improved detection probability compared to the high-user case. This outcome is expected, as mitigating MUI becomes increasingly challenging when more users are present. The average per-user sum rate, illustrated in Figure~\ref{fig:comp_user}~(b), remains approximately constant across different user counts. This observation indicates that the optimization framework effectively maintains a consistent communication rate per user, regardless of system load.

\subsection{Performance Evaluation in a Dynamic Sensing Environment}
The analyses presented in the previous sections were confined to a stationary sensing environment. We now extend the evaluation to a dynamic scenario to investigate system behavior under time-varying conditions. In this case, three targets are considered throughout the simulation, as summarized in Table~\ref{tab:scene2}. Initially, only Target~1 is present. 
\begin{table}[htbp!]
\centering
\caption{Dynamic Scenario}
\label{tab:scene2}
\renewcommand{\arraystretch}{1.2} 
\setlength{\tabcolsep}{8pt} 
\begin{tabular}{c c c c c c c}
\toprule
\textbf{Target} & \multicolumn{6}{c}{\textbf{Time Interval}} \\
\cmidrule(lr){2-7}
 & \textbf{[1, 50]} & \textbf{[51, 100]} & \textbf{[101, 110]} & \textbf{[111, 120]}& \textbf{[121, 130]}& \textbf{[131, 140]}\\
\midrule
\multirow{2}{*}{1} 
  & (-0.4, -0.4)\footnotemark[1] & (-0.4, -0.4) & (-0.4, -0.4) & (-0.4, -0.4) & (-0.4, -0.4) & (-0.4, -0.4)  \\
  & \makecell{$-30$ dB\footnotemark[2]} & \makecell{$-30$ dB} &  \makecell{$-31$ dB} & \makecell{$-32$ dB} & \makecell{$-33$ dB} & \makecell{$-34$ dB} \\
\midrule
\multirow{2}{*}{2} 
  & (0, 0) & Absent & Absent & Absent & Absent & Absent\\
  &  \makecell{$-25$ dB} & -- & --& --& --& --\\
\midrule
  \multirow{2}{*}{3} 
  & Absent & (0.3, 0.1) & (0.3, 0.1) & (0.3, 0.1) & (0.3, 0.1) & (0.3, 0.1) \\
  & -- & \makecell{$-30$ dB} & \makecell{$-30$ dB} & \makecell{$-30$ dB} &\makecell{$-30$ dB} &\makecell{$-30$ dB}\\
\bottomrule
\end{tabular}

\footnotetext[1]{Target’s position.}
\footnotetext[2]{Target’s SNR.}
\end{table}
After 50 time steps, Target~1 disappears and Target~2 emerges with an SNR of $-30$~dB. Subsequently, beginning at time step~100, the SNR of Target~2 gradually decreases. The number of communication users is maintained at $K = 48$, and the following results are analyzed for multiple sensing–communication trade-off factors.

\begin{figure}[htbp!]
    \centering
    \includegraphics[width=1\textwidth]{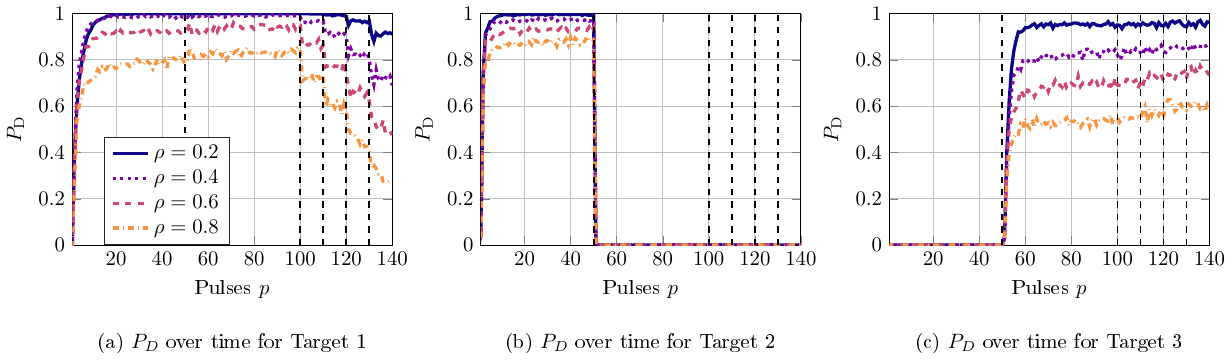}
    \caption{Probability of detection over pulses for Target 1,2 and 3 with $K = 48$ and varying $\rho$}
    \label{fig:dym_prob}
\end{figure}

The evolution of the detection probability over time for all three targets is depicted in Figure~\ref{fig:dym_prob}. The vertical black dashed lines indicate the time instances at which the environment undergoes changes. It can be observed that, for Target~1, the detection probability progressively decreases as $\rho$ increases and the corresponding SNR diminishes. Moreover, the RL-based approach exhibits strong performance in detecting targets that appear at fixed spatial locations. 

\begin{figure}[htbp!]
    \centering
    \includegraphics[width=0.5\textwidth]{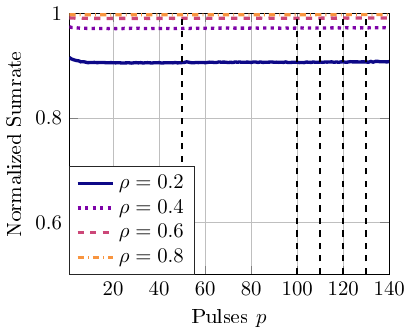}
    \caption{Sum rate over time with $K = 48$ and transmit SNR of $12$~dB and varying $\rho$}
    \label{fig:dym_sumrate}
\end{figure}

The corresponding average sum rate over time is presented in Figure~\ref{fig:dym_sumrate}. The results show that the sum rate remains largely stable, except for an initial transient phase during which the RL agent is still adapting to the target positions, resulting in a slightly elevated sum rate. 

\subsection{Impact of Sequential Target Appearance on Communication Performance}
\begin{figure}[htbp!]
    \centering
    \includegraphics[width=0.5\textwidth]{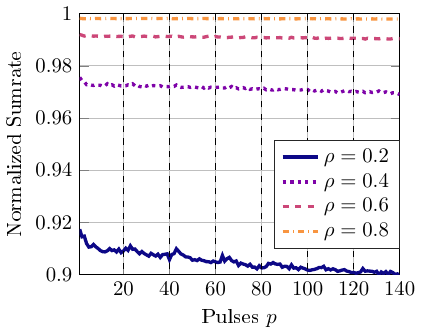}
    \caption{Sum rate over time for $K = 48$ and transmit SNR of $12$~dB, with varying $\rho$ values in a dynamic scenario where a new target appears every 20 time steps.}
    \label{fig:dym_sumrate_add}
\end{figure}
The absence of variation in the sum rate observed in the previous subsection, despite the dynamic environment, can be attributed to the constant number of detected targets in that scenario. To further investigate the system, we now consider a dynamic setting in which new targets are sequentially appearing over time. Specifically, a new target appears every 20 time steps, beginning with a single target and reaching a total of seven targets after 120 time steps. As before, the performance is compared across different trade-off factors $\rho$, with the number of communication users fixed at $K = 48$.

For a trade-off factor of $\rho = 0.2$, Figure~\ref{fig:dym_sumrate_add} clearly shows that the sum rate decreases each time a new target is introduced. This trend is intuitive, as the addition of more targets increases the number of beams required for sensing, thereby reducing the optimization’s ability to effectively suppress MUI when sensing performance is prioritized. A similar decline in the sum rate is observed for $\rho = 0.4$. In contrast, for more communication-oriented configurations with $\rho = 0.6$ and $\rho = 0.8$, this effect is largely negligible.

\section{Conclusions}\label{sec4}
This paper presented an RL-aided cognitive ISAC framework for massive MIMO systems employing a UPA configuration. The proposed approach integrates reinforcement learning into the cognitive radar perception–action cycle to enable adaptive waveform optimization under uncertain and dynamic environments. A Wald-type detector was utilized for robust target detection in non-Gaussian clutter, while a SARSA-based RL agent continuously refined its decision policy to maximize detection performance without prior environmental knowledge.

The proposed design jointly optimized radar sensing and downlink communication by formulating a unified waveform optimization problem that balances the conflicting objectives of sensing accuracy and communication throughput. Through an analytically derived closed-form solution, the framework provided a flexible and computationally efficient method to adjust this trade-off via a single weighting factor. As a result, the system achieved an adaptive equilibrium between multi-user interference suppression and beampattern fidelity.

Comprehensive simulations confirmed that the RL-driven ISAC scheme consistently outperforms both orthogonal and non-learning adaptive baselines, particularly in low-SNR and dynamically varying environments. The results demonstrated that the cognitive learning mechanism enhances detection probability while maintaining stable and competitive communication rates. Furthermore, the study revealed that appropriate trade-off tuning enables the system to maintain balanced performance across varying user densities and channel conditions.

Overall, the presented framework establishes a robust and adaptive foundation for intelligent ISAC operation in next-generation wireless networks. By coupling RL-based learning with analytical waveform optimization, it demonstrates the feasibility of self-optimizing radar–communication coexistence without explicit environmental modeling.

\begin{appendices}
\section{Robust Disturbance Covariance Estimation}\label{secA1}

In general, the disturbance covariance matrix $\mathbf{\Gamma}$ is unknown. In the context of massive MIMO radar~\cite{fortunati2020Wald}, however, it can be estimated directly from the \emph{single} snapshot under test by means of robust techniques for dependent data. The key assumption required for consistent estimation is stated below.

\begin{assumption}\label{assump:dist}
Let $c$ denote a discrete-time, complex circular, possibly non-Gaussian random process $\{c_n, \forall n\}$ whose autocorrelation function decays polynomially. Formally,
\begin{equation}
    r_c[m] \triangleq \mathbb{E}\{c_n\,c_{n-m}^*\} = \mathcal{O}\!\bigl(|m|^{-\gamma}\bigr),
    \quad m \in \mathbb{Z}, \quad \gamma > \frac{\epsilon}{\epsilon - 1},\; \epsilon > 1.
\end{equation}
\end{assumption}

Assumption~\ref{assump:dist} is general enough to capture a wide range of disturbance models, including autoregressive–moving-average (ARMA) processes (possibly non-Gaussian) and compound Gaussian models. Moreover, it extends to higher-dimensional cases (e.g., 2D AR fields), provided that stationarity and polynomial correlation decay hold and the process is suitably vectorized.

\subsection{Sample Covariance Estimation}
Under Assumption~\ref{assump:dist}, the disturbance covariance matrix can be consistently estimated by truncating the empirical covariance at a finite lag~$l$. The entrywise estimator~\cite[Remark~1]{fortunati2020Wald} is defined as
\begin{equation}
\label{eq:GammaEst}
\bigl[\hat{\Gamma}_l\bigr]_{i,j} =
\begin{cases}
\hat{c}_i\,\hat{c}_j^{*}, & j-i \leq l,\\[3pt]
\hat{c}_i^{*}\,\hat{c}_j, & i-j \leq l,\\[3pt]
0, & |i-j| > l,
\end{cases}
\end{equation}
where $\hat{c}_n = y_n - \hat{\alpha}\,h_n$ denotes the residual disturbance after signal subtraction. The truncation parameter $l$ must grow more slowly than $N^{1/3}$ to guarantee consistency~\cite{white1984NLRegression,fortunati2020Wald}, where $N = N_\mathrm{t} N_\mathrm{r}$ is the number of \emph{virtual} spatial antenna channels.

\subsection{Asymptotic Distribution of the Test Statistic}
If Assumption~\ref{assump:dist} holds, Theorem~3 in~\cite{fortunati2020Wald} establishes the following asymptotic results for the Wald-type test statistic $\Lambda_{p,m}$:
\begin{enumerate}
\item[(i)] Under the null hypothesis $\mathcal{H}_0$,
\begin{equation}
\Lambda_{p,m}\bigl(\mathbf{y}_{p,m} \,\big|\,\mathcal{H}_0\bigr)
\;\;\overset{d}{\underset{N\rightarrow \infty}{\sim}}\;\;
\chi^2_{2}(0).
\end{equation}
Thus, the false-alarm rate is constant false alarm rate (CFAR) in the large-system regime, even without secondary data.
\item[(ii)] Under the alternative hypothesis $\mathcal{H}_1$,
\begin{equation}
\Lambda_{p,m}\bigl(\mathbf{y}_{p,m} \,\big|\,\mathcal{H}_1\bigr)
\;\;\overset{d}{\underset{N\rightarrow \infty}{\sim}}\;\;
\chi^2_{2}\!\bigl(\zeta\bigr),
\end{equation}
with non-centrality parameter
\begin{equation}
\zeta = 2\,|\alpha|^{2}\;\frac{\|\mathbf{h}\|^{4}}{\mathbf{h}^{H}\,\mathbf{\Gamma}\,\mathbf{h}}.
\end{equation}
\end{enumerate}

\subsection{Detection Probability in the Large-System Regime}
In the asymptotic regime ($N \to \infty$), the probability of detection at threshold $\eta$ converges to
\begin{equation}
  P_{D}(\eta) \;\underset{N\rightarrow\infty}{\longrightarrow}\;
  Q_{1}\!\bigl(\sqrt{\zeta},\,\sqrt{\eta}\bigr),
\end{equation}
where $Q_{1}(\cdot,\cdot)$ denotes the first-order Marcum $Q$-function~\cite{nuttall1974qfunction}. Once the nominal false-alarm probability $P_{\mathrm{FA}}$ is fixed, the threshold $\eta$ follows accordingly, and $P_D$ becomes a closed-form function of $\zeta$. Importantly, $\zeta$ depends on the disturbance only through the ratio $\|\mathbf{h}\|^{4}/(\mathbf{h}^{H}\mathbf{\Gamma}\mathbf{h})$. Hence, it is sufficient that $\hat{\mathbf{\Gamma}}_l$ consistently estimates $\mathbf{\Gamma}$ in this ratio, even if the exact distribution of the disturbance remains unknown.

\subsection{Discussion}
By leveraging robust estimation theory for dependent observations~\cite{fortunati2023asymp} and the large antenna dimension, the Wald-type detector achieves invariance to the disturbance distribution. Consequently, it enjoys a strong CFAR property without requiring secondary (training) snapshots from homogeneous ranges.

\section{Solution for the trade-off design optimization problem} \label{appB}
To solve $\mathcal{P}_{0.1}$ a low-complexity algorithm is derived in~\cite{liu2018DFRC} that achieves the global optimum in the following. Let us further expand the objective function of in $\mathcal{P}_{0.1}$ as
\begin{align}
\|\mathbf{A}\mathbf{X}-\mathbf{B}\|_F^2
&= \mathrm{tr}\!\big((\mathbf{A}\mathbf{X}-\mathbf{B})^H(\mathbf{A}\mathbf{X}-\mathbf{B})\big) \notag \\
&= \mathrm{tr}\!\big(\mathbf{X}^H\mathbf{A}^H\mathbf{A}\mathbf{X}\big)
   - \mathrm{tr}\!\big(\mathbf{X}^H\mathbf{A}^H\mathbf{B}\big)
   - \mathrm{tr}\!\big(\mathbf{B}^H\mathbf{A}\mathbf{X}\big)
   + \mathrm{tr}\!\big(\mathbf{B}^H\mathbf{B}\big).
\end{align}
Defining $\mathbf{Q}=\mathbf{A}^H\mathbf{A}$ and $\mathbf{G}=\mathbf{A}^H\mathbf{B}$, the problem can be rewritten as
\begin{align}
\min_{\mathbf{X}}\;\; & \mathrm{tr}\!\big(\mathbf{X}^H\mathbf{Q}\mathbf{X}\big) - 2\,\operatorname{Re}\!\big(\mathrm{tr}(\mathbf{X}^H\mathbf{G})\big)
\label{eq:trs-matrix-obj}
\\
\text{s.t.}\;\; & \|\mathbf{X}\|_F^2 = L P_T .
\end{align}
Since $\mathbf{Q}$ is Hermitian, \eqref{eq:trs-matrix-obj} is the matrix version of the trust-region subproblem (TRS), for which strong duality holds. The Lagrangian is
\begin{align}
\mathcal{L}(\mathbf{X},\lambda)
= \mathrm{tr}\!\big(\mathbf{X}^H\mathbf{Q}\mathbf{X}\big)
  - 2\,\operatorname{Re}\!\big(\mathrm{tr}(\mathbf{X}^H\mathbf{G})\big)
  + \lambda\big(\|\mathbf{X}\|_F^2 - L P_T\big),
\end{align}
where $\lambda$ is the dual variable. Let $\mathbf{X}_{\mathrm{opt}}$ and $\lambda_{\mathrm{opt}}$ be primal/dual optima. The optimality conditions are
\begin{subequations}
\begin{align}
\nabla_{\mathbf{X}}\mathcal{L}(\mathbf{X}_{\mathrm{opt}},\lambda_{\mathrm{opt}}) &= 2(\mathbf{Q}+\lambda_{\mathrm{opt}}\mathbf{I}_N)\mathbf{X}_{\mathrm{opt}} - 2\mathbf{G} = \mathbf{0}, \\
\|\mathbf{X}_{\mathrm{opt}}\|_F^2 &= L P_T,\label{eq:lag2}\\
\mathbf{Q}+\lambda_{\mathrm{opt}}\mathbf{I}_N &\succeq \mathbf{0}\label{eq:lag3},
\end{align}
\end{subequations}
which yield
\begin{align}
\mathbf{X}_{\mathrm{opt}} = (\mathbf{Q}+\lambda_{\mathrm{opt}}\mathbf{I}_N)^{\dagger}\mathbf{G}.\label{eq:bas1}
\end{align}
Using~\eqref{eq:lag2} and~\eqref{eq:lag3} we have
\begin{align}
\Big\|(\mathbf{Q}+\lambda_{\mathrm{opt}}\mathbf{I}_N)^{\dagger}\mathbf{G}\Big\|_F^2
= \Big\|\mathbf{V}(\boldsymbol{\Lambda}+\lambda_{\mathrm{opt}}\mathbf{I}_N)^{-1}\mathbf{V}^H\mathbf{G}\Big\|_F^2
= L P_T,\quad \lambda_{\mathrm{opt}}\ge -\lambda_{\min},
\end{align}
where $\mathbf{Q}=\mathbf{V}\boldsymbol{\Lambda}\mathbf{V}^H$ is the eigen-decomposition with $\mathbf{V}$ unitary and diagonal $\boldsymbol{\Lambda}$, and $\lambda_{\min}$ is the minimum eigenvalue of $\mathbf{Q}$. Define
\begin{align}
P(\lambda)
&= \Big\|\mathbf{V}(\boldsymbol{\Lambda}+\lambda\mathbf{I}_N)^{-1}\mathbf{V}^H\mathbf{G}\Big\|_F^2
= \sum_{i=1}^{N}\sum_{j=1}^{L} \frac{\big([\mathbf{V}^H\mathbf{G}]_{i,j}\big)^2}{(\lambda+\lambda_i)^2},
\end{align}
which is strictly decreasing and convex for $\lambda\ge -\lambda_{\min}$. Hence $\lambda_{\mathrm{opt}}$ can be found efficiently by a 1-D line search (e.g., Golden-section), after which $\mathbf{X}_{\mathrm{opt}}$ follows from~\eqref{eq:bas1}. For clarity, we summarize the above approach in Algorithm~\ref{alg:low-complexity}.

\begin{algorithm}[t]
\caption{Low-complexity Algorithm}
\label{alg:low-complexity}
\textbf{Input:} $\mathbf{H},\mathbf{S},\mathbf{X}_0$, weight $0\le \rho \le 1$, $P_T$.\\
\textbf{Output:} Global minimizer $\mathbf{X}_{\mathrm{opt}}$.
\begin{enumerate}
\item Form $\mathbf{A},\mathbf{B}$ as in (18) and compute $\mathbf{Q}=\mathbf{A}^H\mathbf{A}$, $\mathbf{G}=\mathbf{A}^H\mathbf{B}$.
\item Eigendecompose $\mathbf{Q}=\mathbf{V}\boldsymbol{\Lambda}\mathbf{V}^H$ and set the search interval $[\,-\lambda_{\min},\,b\,]$ with $b\ge 0$.
\item Find $\lambda_{\mathrm{opt}}$ via line search (e.g., Golden-section).
\item Set $\mathbf{X}_{\mathrm{opt}} = (\mathbf{Q}+\lambda_{\mathrm{opt}}\mathbf{I}_N)^{\dagger}\mathbf{G}$.
\end{enumerate}
\end{algorithm}

\section{Solution for the radar waveform optimization problem} \label{appC}
Given a target covariance matrix $\mathbf{R}_d$ corresponding to a well-designed MIMO radar beampattern, the MUI minimization problem can be formulated as
\begin{align}
\min_{\mathbf{X}_0} \;\; & \left\|\mathbf{H}\mathbf{X}_0-\mathbf{S}\right\|_F^2 \nonumber\\
\text{s.t.} \;\; & \tfrac{1}{L}\mathbf{X}_0\mathbf{X}_0^H=\mathbf{R}_d.
\end{align}
 The matrix $\mathbf{R}_d$ is Hermitian and positive semidefinite.  

We consider the Cholesky factorization
\begin{align}
\mathbf{R}_d = \mathbf{F}\mathbf{F}^H ,
\end{align}
where $\mathbf{F}\in\mathbb{C}^{N\times N}$ is lower triangular. Without loss of generality, we assume $\mathbf{R}_d$ is positive definite, which guarantees that $\mathbf{F}$ is invertible. Substituting this decomposition into the constraint yields
\begin{align}
\tfrac{1}{L}\,\mathbf{F}^{-1}\mathbf{X}_0\mathbf{X}_0^H\mathbf{F}^{-H} = \mathbf{I}_N .
\end{align}

Defining the normalized variable
\begin{align}
\widetilde{\mathbf{X}}_0 \triangleq \sqrt{\tfrac{1}{L}}\,\mathbf{F}^{-1}\mathbf{X}_0,
\end{align}
the problem reduces to
\begin{align}
\min_{\widetilde{\mathbf{X}}_0} \;\; & \left\|\sqrt{L}\,\mathbf{H}\mathbf{F}\,\widetilde{\mathbf{X}}_0-\mathbf{S}\right\|_F^2 \nonumber\\
\text{s.t.} \;\; & \widetilde{\mathbf{X}}_0\widetilde{\mathbf{X}}_0^H=\mathbf{I}_N .
\end{align}
This is recognized as an orthogonal Procrustes problem (OPP). Its globally optimal solution is
\begin{align}
\widetilde{\mathbf{X}}_0 = \widetilde{\mathbf{U}}\,\mathbf{I}_{N\times L}\,\widetilde{\mathbf{V}}^H ,
\end{align}
where$\widetilde{\mathbf{U}}\,\widetilde{\boldsymbol{\Sigma}}\,\widetilde{\mathbf{V}}^H = \mathbf{F}^H\mathbf{H}^H\mathbf{S}$ is the singular value decomposition (SVD) of $\mathbf{F}^H\mathbf{H}^H\mathbf{S}$.  

Finally, substituting back, the optimal transmit waveform matrix is obtained as
\begin{align}
\mathbf{X}_0 = \sqrt{L}\,\mathbf{F}\,\widetilde{\mathbf{U}}\,\mathbf{I}_{N\times L}\,\widetilde{\mathbf{V}}^H .
\end{align}

\end{appendices}
\section*{Abbreviations}
\renewcommand*{\arraystretch}{1.37}
\begin{longtable}{@{}l @{\hspace{5mm}} l }

2D                 &  Two-dimensional\\
6G                 &  Sixth-generation\\
AR                 &  Autoregressive\\
ARMA               &  Autoregressive–moving-average\\
AWGN               &  Additive white Gaussian noise\\
BS                 &  Base station\\
CFAR               &  Constant false alarm rate\\
CR                 &  Cognitive Radar\\
GLRT               &  Generalized likelihood ratio test\\
ISAC               &  Integrated sensing and communication\\
MIMO               &  Multiple-input and multiple-output\\
MUI                &  Multi-user interference\\
NRL                &  Non reinforcement learning \\
PSD                &  Power spectral density \\
QCQP               &  Quadratically constrained quadratic program\\
RL                 &  Reinforcement learning\\
SARSA              &  State–action–reward–state–action\\
SDP                &  Semidefinite Program\\
SDR                &  Semidefinite Relaxation\\
SER                &  Symbol error rate\\
SINR               &  Signal-to-interference-plus-noise ratio\\
SNR                &  Signal-to-noise ratio\\
STAP               &  Space–time adaptive processing\\
SVD                &  Singular value decomposition\\
TRS                &  Trust-region subproblem\\
ULA                &  Uniform linear array\\
UPA                &  Uniform planar array\\
QPSK               &  Quadrature phase-shift keying\\

\end{longtable}

\section*{Declaration}
\subsection*{Availability of data and materials}
The datasets used and analysed during the current study are available from the corresponding author on reasonable request.

\subsection*{Competing interests}
The authors declare that they have no competing interests.
\subsection*{Funding}
This work was supported by the German Research Foundation
(“Deutsche Forschungsgemeinschaft”) (DFG) under Project–ID 287022738
TRR 196 for Project S03.
\subsection*{Authors' contributions}
AU conceived the study, implemented it, conducted the analyses, and wrote the manuscript. All authors were involved in idea development and manuscript review, and all approved the final manuscript.
\subsection*{Acknowledgements}
Not applicable.

\clearpage
\bibliography{sn-bibliography}

\end{document}